# SG-NNP: Species-separated Gaussian Neural Network Potential with Linear Elemental Scaling and Optimized Dimensions for Single and Multi-component Materials

Ji Wei Yoon[1,2*], Bangjian Zhou[1], J Senthilnath[1]

[1]*Institute for Infocomm Research (I²R), Agency for Science, Technology and Research (A\*STAR), 1 Fusionopolis Way, #21-01, Connexis South Tower, Singapore 138632, Republic of Singapore*

[2]*Department of Materials Science and Engineering, University of California, Berkeley, CA 94720, USA.*

*Corresponding author

**Abstract**

Accurate simulations of materials at long-time and large-length scales have increasingly been enabled by Machine-learned Interatomic Potentials (MLIPs). There have been increasing interest on improving the robustness of such models. To this end, we engineer a novel set of Gaussian-type descriptors that scale linearly with the number of atoms, reduce informational degeneracy for multi-component atomic environments and apply them in Species-separated Gaussian Neural Network Potentials (SG-NNPs). The robustness of our method was tested by analyzing the impact of various design choices and hyperparameters on Molybdenum (Mo) SG-NNPs' performance during training and inference/simulation. With less dimensions, SG-NNPs are shown to have superior atomic forces and total energy predictions than other traditional and ML *descriptor-based* interatomic potentials on diverse set of materials – Ni, Cu, Li, Mo, Si, Ge, NiMo, Li$_3$N and NbMoTaW. From the obtained results we can observe that the proposed method improves the performance of atomic descriptors of complex environments with multiple species.

## 1. Introduction

Modern machine-learned interatomic potentials (MLIPs) have unlocked a variety of simulations that have, till recently, been impossible due to a lack of accurate and efficient potentials. By parameterizing the potential energy surface (PES) using local environmental descriptors and coupling them with ML models, MLIPs are enabling longer time and larger length scale simulations[1–3]. In doing so, MLIPs are providing insights into a wide range of phenomena in complex materials systems, for example, catalysts[4,5], magnetic systems[6], and multiple principal element alloys[7,8].

There has been continued interest in finding the right combination of ML techniques and atomic fingerprints/descriptors for materials containing many elements (multi-component systems). These materials exhibit complex behaviors and are rich ground to mine for useful applications. However, it remains challenging to model them accurately and efficiently. This is partly due to the vast configurational space that must be represented accurately. Moreover, there have been concerns about the resolution[9–11] and completeness[12,13] of popular descriptors like SOAP[14] and ACSF[15], which are applied in SNAP[16,17] and HDNNP[18] potentials, respectively. Also, the dimensionality of such many-body descriptors scales poorly with the number of elements. This issue was addressed for the SOAP descriptor in a recent study[19], where the dimensionality scaling is linearized, from quadratic, with respect to the number of elements.

Despite the known deficiencies of SOAP and ACSF, they have been applied to various degrees of success in a variety of materials and conditions and are actively being extended to encompass more physics. Recently, SNAPs have been developed for W-ZrC in extreme reactor conditions[20] and C at extreme pressures[21]. HDNNPs have been extended to include spins to model magnetic interactions[6] and charge transfer for more accurate electrostatics[22]. We continue this line of work to extend the performance of such descriptors by attempting to reduce informational degeneracy and making their dimensionality scale linearly with the number of atoms.

This work is motivated by the high accuracy of the Radial and Angular Distribution Functions (RDF and ADF) used by Artrith et al.[23], which fitted the energies of a DFT dataset with 11 chemical species down to a root mean squared (RMSE) error for the forces below 5 meV/atom. The novelty of our approach lies in the delineation of positional information and elemental identity in modified RDF and ADF to generate salient descriptors. Moreover, we base our use of the Gaussian convolutional kernel on a signal processing theorem that guarantees a unique representation of the modified RDF and ADF. As far as we know, the uniqueness theorem has not been adopted before. Going beyond that, due to the known sensitivity of Neural Networks to their hyperparameters, we conduct comprehensive investigations of the impact of various design choices and hyperparameters on model loss. For hyperparameters related to the Gaussian kernels, we optimize them with a gradient-free Response Surface Methodology and efficient Latin Hypercube sampling to maximize model accuracy while keeping the number of descriptors small. It is shown that our descriptors only scale linearly with the expected number of neighbor atoms in the cutoff radius and not the number of elements.

The recent availability of public datasets on single and multi-component materials has allowed systematic comparisons of newly developed descriptors and methods with other well-established ones. In this work, we leverage the datasets[7,24–26] by the Materials Virtual Lab at UC San Diego and show that our method is able to produce accurate atomic forces and total energies predictions with our bespoke descriptors with a lower number of dimensions than SNAP and HDNNP for a diverse set of materials: Ni, Cu, Li, Mo, Si, Ge, NiMo, $Li_3N$ and NbMoTaW..

To ensure efficiency, we parallelize the descriptor engineering process by leveraging a concurrent multi-nodal training process using DASK[27] and MPI. Also, we engineer an efficient implementation of SG-NNPs in the OpenKIM framework[28], which is widely portable to different molecular dynamics calculators, e.g. the widely-used LAMMPS[29].

In summary, the key contributions of this work are as follows:

1. To engineer a novel set of descriptors, based on mathematical uniqueness guaranteed by a signal processing theorem, for multicomponent materials that reduces informational degeneracy and scales only linearly with the number of atoms

2. To use Latin Hypercube Sampling and Response Surface Methodology to tune the hyperparameters for the Gaussian kernels to generate descriptors that maximize accuracy while keeping the number of dimensions low
3. To conduct comprehensive investigations of all other design choices and hyperparameters on model loss
4. To implement parallel descriptor engineering using DASK and MPI
5. To implement efficient OpenKIM implementations of our potentials
6. To comprehensively compare our method with other existing MLIPs on datasets of diverse materials

## 2. Results and Discussion

In this section, we focus on the tuning and fitting of an SG-NNP using a DFT Molybdenum (Mo) dataset. We conduct comprehensive hyperparameter tuning and investigate the effects of various design choices. We use our Mo SG-NNP to investigate Mo material properties and compare that with a Spectral Neighbor Analysis SNAP potential (SNAP), which was fit to the same dataset, an Embedded Atom Method (EAM) potential, a Modified Embedded Atom Method (MEAM) potential and a Morse potential. Finally, we also fit our SG-NNP to a comprehensive set of single and multicomponent materials and compare it with other modern MLIPs.

### 2.1 Species-separated Gaussian Representations of Multi-component Atomic Environments

We elaborate on the theoretical construction that allows us to describe the local atomic environment. Our features are different from the descriptors introduced by Behler et al. [30,31], who used a set of pre-determined hyperparameters for their Gaussian kernels. We relax this condition and optimize over a range of possible widths and means of the Gaussians. Also, we use the cosine of the angle for our angular descriptors instead of a transformed factor composed of the cosine of the angle subtended by three atoms. Furthermore, the Gaussians in the angular descriptor in Ref. [30] use the interatomic separation distance while we employ an angular descriptor with the cosine angle.

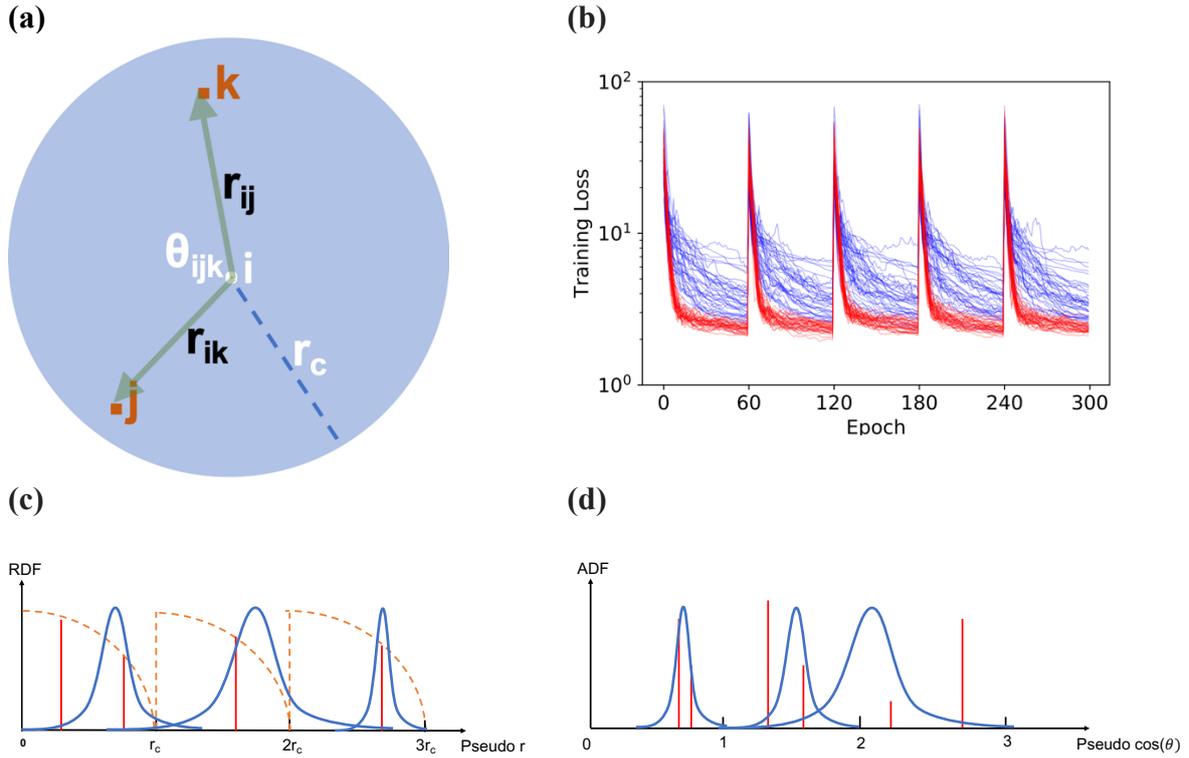

Figure 1 Species-separated Gaussian Representations. (a) A three-atom atomic environment of central atom i and neighbour atoms j and k with cut-off radius $r_c$. (b) The training loss versus epoch graph of a two-stage five-fold cross-validation training run for the Molybdenum dataset. A second-order LBFGS optimizer with line search and strong Wolfe conditions was used in the training process shown in the figure. First-stage and second-stage training losses are in blue and red, respectively. (c) The Radial Distribution Function (RDF) of a four-atom neighborhood with three species of particles represented by three Dirac deltas. The cut-off function is shown as the dotted line. (d) The Angular Distribution Function (ADF) is associated with the same four-atom neighborhood with $\binom{3}{2}$ = 3 distinct pair types and $\binom{4}{2}$ = 6 Dirac deltas.

Figure 1 (a) depicts the atomic environment of the central atom *i* with cut-off radius $r_c$. In this configuration, there are two neighbor atoms to atom *i*, namely *j* and *k*. They are separated from *i* by distances $r_{ij}$ and $r_{ik}$, respectively. This three-atom configuration is also described by the angular separation $\theta_{ijk}$, which is the angle subtended at *i* by *j* and *k*.

In general, the atomic configuration centered on atom *i* is made up of all the atoms contained within the cut-off radius $r_c$ from atom *i*. In our methodology, the total energy of the system is expressed as a sum of atomic energies ($E_i$) dependent on the local atomic configuration:

$$E_{total} = \sum_{i}^{N} E_i \qquad (1)$$

$$E_i = F_{NN}\left(\underline{\alpha_i}, \underline{\gamma_i}\right) \qquad (2)$$

where $\underline{\alpha_i}, \underline{\gamma_i}$ are the descriptors describing the local atomic environment and are the inputs to the neural network $F_{NN}$. Specifically, $\alpha_i$ is referred to as a Radial Gaussian Coefficient (RGC) and $\gamma_i$ is referred to as an Angular Gaussian Coefficient (AGC). These variables are simply the projection of the Radial Distribution Function (RDF) and Angular Distribution Function (ADF) onto Gaussians placed in their respective domains.

We define the RDF for atom $i$'s environment as:

$$RDF_i(r) = \sum_{j \in cutoff} \delta\left(r - (\tau_j + r_{ij})\right) f_c(r_{ij}) \qquad (3)$$

where $j$ is the neighbor of atom $i$, $\tau_j$ is a species-dependent shift factor, $r_{ij}$ is the separation distance between $i$ and $j$, $\delta$ is the Dirac delta function and $f_c$ is the cutoff function defined as:

$$f_c(r_{ij}) = \begin{cases} 0.5 * \left(\cos\left(\frac{\pi r_{ij}}{r_c}\right) + 1\right), & 0 \leq r_{ij} \leq r_c \\ 0 & , r_{ij} > r_c \end{cases} \qquad (4)$$

The cut-off function reduces the $\delta$'s associated with an atom smoothly to zero height as the atom moves out of the cut-off sphere. For central atom $i$, the associated radial descriptor, which we refer to as RGC, for a Gaussian centered on $\mu_v$ with standard deviation $\sigma_v$ is:

$$\alpha_{iv} = \sum_{j \in cutoff} e^{-\frac{(\tau_j + r_{ij} - \mu_v)^2}{2\sigma_v^2}} f_c(r_{ij}) \qquad (5)$$

We define the ADF for atom i's environment as:

$$ADF_i(\theta) = \sum_{j,k \in cutoff} \delta\left(\theta - (\omega_{jk} + \cos(\theta_{ijk}))\right) f_c(r_{ij}) f_c(r_{ik}) \qquad (6)$$

where $\omega_{jk}$ is a pair-dependent shift factor, $\theta_{ijk}$ is the angle subtended at atom $i$ between atom $j$ and $k$, $\delta$ is the Dirac delta function and $f_c$ is the cut-off function defined before. We wish to emphasize that $\theta$, $\omega_{jk}$ and $\cos(\theta_{ijk})$ are all unitless. The argument $\theta$ of the function is referred to as "pseudo $\cos(\theta)$" because it refers to specific values in the space that contains Dirac deltas placed at the values of shifted $\cos(\theta_{ijk})$ (i.e. $\omega_{jk} + \cos(\theta_{ijk})$ ).

For the central atom $i$, the associated angular descriptor, which we refer to as the Angular Gaussian Coefficient (AGC), for a Gaussian centered on $\mu_v$ with standard deviation $\sigma_v$ is,

$$\gamma_{iv} = \sum_{j,k \in cutoff} e^{-\frac{(\omega_{jk} + \cos(\theta_{ijk}) - \mu_v)^2}{2\sigma_v^2}} f_c(r_{ij}) f_c(r_{ik}) \qquad (7)$$

Besides the RGC and AGC, for the purpose of fitting to data including forces, and to compute forces for atomistic simulations, we require the derivatives of these functions with respect to atomic coordinates. These expressions are provided in the Supplementary Equations, along with a detailed derivation of the quantities defined above.

The motivation for representing the RDF and ADF with Gaussians comes from Theorem 6 of Ref. [32], from the signal processing literature:

*"Given a finite stream of K weighted Diracs and a Gaussian sampling kernel $h_\sigma(t) = exp(-t^2/2\sigma^2)$. If N≥2K, then the N sample values $y_n = \langle h_\sigma(t - nT), x(t)\rangle$ are sufficient to reconstruct the signal."*

The theorem gives the theoretical guarantee of the uniqueness of the Gaussian Coefficient representation of the RDF and ADF under the condition that the number of Gaussians must be at least twice the number of Dirac delta functions (δ). This is because there are two pieces of information per δ – its position and height. We note that the Gaussian kernels are equally spaced apart and have the same width in the theorem. This restriction is loosened in the Gaussian hyperparameter search as it may be better to focus on specific dense parts of the ADF or RDF. We further discuss issues related to the uniqueness and complexity of the representations in Supplementary Section 1.

For systems with a single species, like the Mo elemental systems, there is only a single interval in both the RDF and ADF. The lengths of the RDF and ADF interval are $r_c$ and 1, respectively. There is only one $\tau_j$ for j=Mo and one $\omega_{jk}$ for j and k = Mo and they are chosen to be 0 Å and 0, respectively.

To represent the RDF and ADF of systems with more species, we allow $\tau_j$ to take values to separate all the species in RDF and $\omega_{jk}$ to take values to separate all the pairs of species in ADF space. Typically, $\tau_j$ will be chosen to be a multiple of $r_c$ and $\omega_{jk}$ to be a multiple of 1, where the multiple is, $m_{RDF} \in \{0,1,\ldots,N_s\}$ and $m_{ADF} \in \{0,1,\ldots,\binom{N_s}{2}\}$, respectively, and $N_s$ is the number of distinct species. We further discuss the reasons for the design choice in the Methods (section 3).

Each configuration is associated with its own atomic energy, $E_i$, which is a function of descriptors, $C_n = (\alpha_n, \gamma_n)$, through a feed-forward architecture. The whole number *n* runs from 1 to $N_d$, the total number of descriptors. We employ RGC and AGC as the descriptors. The total energy of the system is a simple sum of all atomic energies.

## 2.2 Implementation

In this section, we describe the training workflow to generate a deployment model. We employed the loss function (refer to Methods) to guide the training process. The averaged validation loss over N fold are tracked. The hyperparameters were optimized using a Latin Hypercube method for initialization and a response surface method for optimization. We refer the reader to Supplementary Method 1 for detailed steps in our workflow.

Figure 1(b) depicts the training loss versus index of epochs of a two-stage five-fold cross-validation training run for the Molybdenum dataset with 60 epochs for each separate training run. The red lines are the second-stage runs with those Response Surface optimized hyperparameter values, which has lower training losses than the first-stage runs. The same was observed for testing errors.

Note that a random selection of Gaussian widths and means in the first-stage runs result in training loss that span half an order of magnitude. We observe the same effect size in the testing losses as well. This shows the importance of the Gaussian hyperparameters (i.e. their means and widths) in determining the quality of the fit even for flexible neural networks. The optimization of the hyperparameters is a crucial step to getting high-quality potentials.

## 2.3 Impact of Design Choices and Hyperparameters

In this section, we summarize a comprehensive investigation of the impact of the multitude of design choices and hyperparameters of our training procedure and the NN potential. The loss on a random 20% portion of the Molybdenum dataset is reported. The aim is to find a set of design choices and hyperparameters associated with a level of accuracy within the magnitude of the reported dataset error. As the number of such choices and hyperparameters is large, we highlight a few that the training loss is most sensitive to or pertains to novel elements of our method. The rest of the set is included in Supplementary Discussion 2.

### 2.3.1 First-order Optimizers and Hyperparameters

In this section, we use first-order optimizers to train the NN potential. First-order optimizers only use gradients of the loss function or their approximations to decide the direction and magnitude to change the parameters. The loss function is defined as the sum of the mean absolute error of the energies and forces.

There are two goals: (1) to ascertain the efficacy and limits of the first-order optimizer in training a non-linear interatomic potential with our Gaussian Coefficients and, (2) to investigate the efficacy of first-order mini-batch-based methods, which are less memory-intensive than second-order methods for large datasets.

#### 2.3.1.1 Choice of Initial Learning Rate

The choice of learning rate for training an NN has a strong influence over the rate of convergence as well as the model's performance. We use two methods to determine the initial learning rate that we use in our cyclic learning rate schedule and investigate the effects of the training loss.

We increase the learning rate progressively till the loss is strongly diverging. Supplementary Figure 9 shows the training loss curves, each with a different random choice of Gaussian hyperparameters in the fixed range. We observe that the training losses have a minimum in the domain $[10^{-3}, 10^{-1}]$, with runs at the top of the range diverging and those near the bottom stagnating. Thus, we investigate two options for picking a learning rate: (1) The common minimum - where most losses have their minimum. (2) The lowest minimum - where the losses are minimized.

**(a)**

|  | MAE Loss | Choice of Initial Learning Rate | |
|---|---|---|---|
|  |  | Common Minimum | Lowest Minimum |
| Architecture | MNL | 7.97 | 4.37 |
|  | $3^1$ | 17.06 | 4.97 |

|  | 96¹ | 7.41 | 4.45 |
|---|---|---|---|

**(b)**

|  | \multicolumn{5}{c}{Architecture} |
|---|---|---|---|---|---|
|  | MNL | 3¹ | 96¹ | 3² | 96² |
| **MAE** | 0.005 | 0.001 | 0.00005 | 0.001 | 0.00005 |
| **RMSE** | 0.01 | 0.005 | 0.0005 | 0.005 | 0.0005 |

Table 1 Statistics of training loss as a function of the choice of initial learning rate and model architecture. (a) The lowest MAE losses of training runs with different architectures and different methods of choosing the initial learning rate. The architecture label $z^b$ refers to a model with b non-linear layers and one linear layer, each with z nodes, and MNL refer to the minimally non-linear model. (b) The initial learning rate was determined using the lowest minimum method for different loss functions and architecture for the RMSprop optimizer. The architecture label $z^b$ refers to a model with b non-linear layers and one linear layer, each with z nodes, and MNL refer to the minimally non-linear model.

Table 1(a) shows the lowest mean absolute error (MAE) loss of three architectures and the two definitions of initial learning rate. There is a drastic improvement in loss when the lowest minimum initial learning rate is used. Remarkably, the minimally non-linear model has the lowest loss amongst models that have more parameters.

Given the drastic improvements, we have chosen to use the lowest minimum as the initial learning rate from this section onwards. Table 1(b) lists the lowest minimum initial learning rate determined for the various architectures and loss functions.

### 2.3.1.2 Search Method for Gaussian Means

Due to the large search space for means of the Gaussians, we investigate a few common-sense placement strategies.

| **Radial Gaussian Means** | **Angular Gaussian Means** | **MAE Loss** |
|---|---|---|
| Residual Eating [0,1] | Residual Eating [0,1] | 4.37 |
| Uniform [0,1] | Uniform [0,1] | 4.65 |
| Uniform [0,1] | Uniform [0,0.5] | 4.45 |
| Uniform [0,1] | 0.5 | 4.40 |
| 1 | 0.5 | 7.97 |
| 1 | Uniform [0,1] | 7.02 |
| 1 | Uniform [0,0.5] | 7.53 |

Table 2 The best MAE training losses of one-fold two-stage training runs of the Neural Network model with different methods of selecting the means of Gaussian kernels.

Table 2 lists the best MAE training losses of NN models with different methods of selecting the means of Gaussians. We elaborate more on the methods in Supplementary Method

2. The widths were selected uniformly in a random fashion in the first stage in the range [0.2,1] in units of $r_c$ for RDF coefficients and 1 for ADF coefficients.

The best MAE losses show that the biggest determinant is whether the RGC's means are spread out or not. The difference in losses is around 50% of the lower value. Placing the AGC means at a specific point does not affect the results to the same extent. We believe that the higher degree of redundancy in the informational content of the ADF over the RDF is the reason.

### 2.3.1.3 Mini-batch Size

The choice of mini-batch size affects the quality of the gradients that are used to update the parameters of the model. While the use of the full batch (i.e. all of the datapoints) yields the exact gradient for a choice of the loss function, there is a possibility that the model will get stuck at a suboptimal solution (local minimum). Hence, the stochasticity of mini-batch gradients may be preferred. Moreover, if the dataset is large, the exact gradient may not be computationally tractable. Generally, the best MAE loss occurs at a moderate mini-batch size.

|  | MAE Loss | Mini-batch Size | | | |
|---|---|---|---|---|---|
|  |  | 3 | 9 | 27 | 280 (Full) |
| Architecture | Minimally Non-Linear | 4.37 | 4.53 | 4.78 | 11.45 |
|  | 3[1] | 4.97 | 4.20 | 4.38 | 7.95 |
|  | 96[1] | 4.45 | 6.43 | 3.74 | 5.24 |

Table 3 The lowest MAE losses of training runs with different architectures and different mini-batch sizes.

### 2.3.1.4 Choice of MSE versus MAE

We have focused on MAE to compare our method with a Mo SNAP[33]. As training a Neural Network typically involves optimization of a non-convex loss landscape, a change in loss function may actually make the loss landscape more amenable to optimization and yield better MAE in energies and forces. We compare MAE and root-mean-squared-error (RMSE), which pays stronger attention to outliers.

|  |  | MAE | | | RMSE | | |
|---|---|---|---|---|---|---|---|
|  |  | Loss | MAE in Energies (meV) | MAE in Forces (eV/Å) | Loss | MAE in Energies (meV) | MAE in Forces (eV/Å) |
| Architecture | Minimally Non-Linear | 4.37 | 16 | 0.58 | 5.42 | 19 | 0.52 |
|  | 3[1] | 4.97 | 18 | 0.70 | 5.06 | 16 | 0.60 |
|  | 3[2] | 6.19 | 24 | 0.77 | 5.72 | 16 | 0.80 |
|  | 96 | 4.45 | 22 | 0.54 | 3.60 | 10 | 0.52 |

| | | | | | | |
|---|---|---|---|---|---|---|
| $96^2$ | 6.21 | 25 | 0.70 | 3.98 | 10 | 0.55 |

Table 4 The lowest MAE and RMSE losses, and their associated MAE in energies and forces, of training runs with different architectures and different loss functions. The architecture label $z^b$ refers to a model with b non-linear layers and one linear layer, each with z nodes. The tested loss functions are MAE and RMSE, which are also the reported loss values.

Table 4 shows the result of using the sum of MAE or RMSE in energies and forces as loss functions and their values with different architectures. Larger models trained with RMSE loss function yield better losses than MAE. Therefore, we choose to use RMSE over MAE loss.

The model trained with RMSE loss function and 2 non-linear layers with 96 nodes each has a training MAE in energies of 10 meV/atom and MAE in forces of 0.52 eV/Å. The comparable value by the Mo Spectral Neighbor Analysis Potential(SNAP) model[33] is 8.6 meV and 0.60 eV/Å by our calculations. The lower reported MAE in forces in [33] is reproduced when we take the absolute average of the force error components instead of the force error vectors. Our model uses a cut-off radius of only 4Å while the SNAP model uses 4.62Å. The number of descriptors of our model is 10 while that of SNAP is 30.

### 2.3.2 Second-order Optimizers and Hyperparameters

It has been shown in a previous work [34] that second-order methods are able to train NN potentials to higher levels of accuracy than first-order ones. We seek to minimize the training loss further and so explore specifically the LBFGS optimizer and its hyperparameters. Note that second-order optimizers are typically more compute and memory intensive than first-order ones because they need to calculate and store an exact or approximate Hessian matrix – a comparative study is discussed in Supplementary Discussion 2.2.1. We investigate the impact of using different (1) widths, (2) number of Gaussians, (3) Gaussian width search approach and (4) widths versus depth of neural network in Supplementary Discussion 2.2.(2-5), respectively.

#### 2.3.2.1 Hyperparameters of LBFGS optimizer

We perform a perturbative test by either halving or doubling the hyperparameter. The results in Supplementary Table 17 show that the LBFGS optimizer with line search and strong Wolfe conditions has the lowest loss. Moreover, the algorithm requires no tuning and trains more stably than the one without line search. From this section onwards, we use this particular optimizer.

#### 2.3.2.2 Number of RGC versus AGC

In our investigations above, we have used an equal number of RGC and AGC. Here, we investigate the RMSE Losses of 16 coefficients distributed differently - (Number of RGC, Number of AGC, RMSE Loss): (8, 8, 2.190), (6, 10, 2.105), (3, 13, 5.723). The RDF contains a third of the total informational content of the atomic coordinates. Assuming the ADF contains the other two-third of the information, we argue that the optimal way to distribute 16 Gaussians is roughly 6 for RDF and 10 for ADF, which is observed.

### 2.4 Molybdenum

Here, we discuss the properties of our Mo SG-NNP, the results from the training and hyperparameter optimization process and its predictive performance against previously developed SNAP[33], EAM[35], MEAM[36] and Morse[37] potentials for Mo. We summarize the models' characteristics in Supplementary Table 20. In the interest of space, we report elastic constants, state equation, vacancy formation and migration energies, surface and grain boundary energies, linear thermal expansion coefficient, dislocation core structures, moment frequency of phonon density of states in Supplementary Discussion 5-11 and Supplementary Table 21-23.

### 2.4.1 Computational Efficiency

We optimize the original implementation of our SG-NNP potential to make it roughly 20 times more efficient than a naïve first implementation. An example of the function call graph, generated from Valgrind [38] with gprof2dot [39], that was used to aid our optimization of the implementation is shown in Supplementary Figure 9.

To test the computational efficiency of the Mo potential, we ran a serial simulation with 2000 atoms in a Body Centred Cubic (BCC) lattice for 100 timesteps on a computer with an Intel Xeon E5-2670 v2 processor. We use the OpenKIM [28,40] interface of LAMMPS[41] to access these potentials. We report the number of timestep times the number of atoms the simulators undertook for each second passed in the simulation relative to MEAM: (1) SG-NNP – 0.277, (2) SNAP – 0.655, (3) EAM – 61.2, (4) Morse - 8.85, MEAM – 1.

The time required to evaluate the ML potentials is within an order of magnitude from the previously published state-of-the-art MEAM potential [36]. We expect further work on reducing the redundancy in ADF will bring the speed of SG-NNP potential comparable to the other potentials. The redundancy in ADF is evident for local configurations with many atoms where the number of Dirac deltas (i.e. $n_{neighbors}$ choose 2) is greater than the angular degrees of freedom (i.e. $2*n_{neighbors}$).

### 2.4.2 Forces and Total Energies Predictions

Overall, our SG-NNP model gives the lowest MAE in energies and forces across all the data. SNAP is second in MAE in energies while MEAM is second in MAE in forces. The categorical breakdown of errors shows that the SG-NNP is in the first position for almost all categories except surface and vacancy. In the case of surface and vacancy structures, the absolute differences in errors between the SG-NNP model and the best model are ~ 1 meV/atom for energies and greater than 0.02 eV/Å for forces. The two values are the reported converged DFT errors for the dataset. See Supplementary Discussion 4 for more analyses.

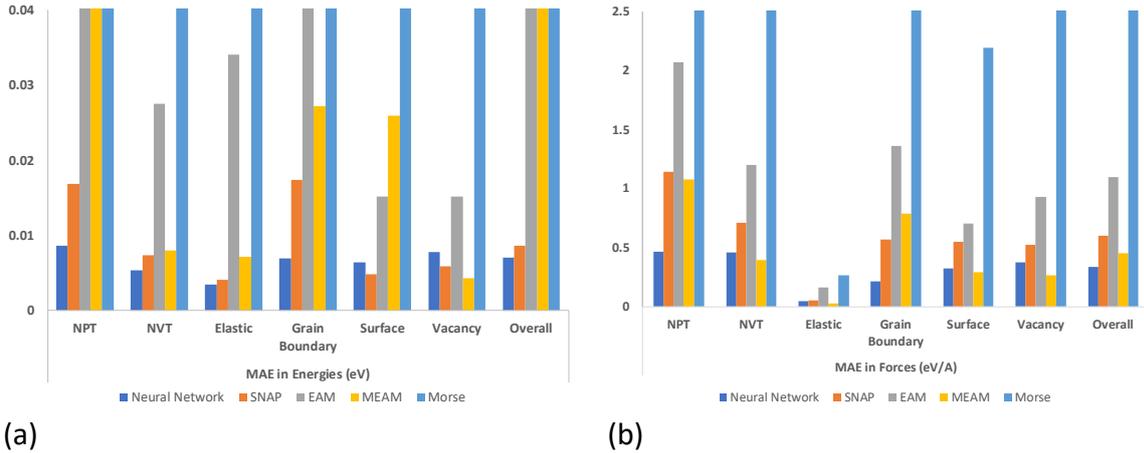

(a) (b)

Figure 2 MAE in energies and forces for different splits of the data. (a) MAE in energies of our deployed Neural Network model (SG-NNP), SNAP, EAM, MEAM and Morse potential, broken down into different non-overlapping subsets of the training data and the whole dataset ('Overall'). (b) The same as (a) but for MAE in forces. Note: Some of the bars that reach the top of the range go beyond the range.

### 2.4.3 Lattice Constants, Cohesive and Formation Energies

We compare the DFT lattice constants, cohesive and formation energies of BCC, Face Centred Cubic (FCC), diamond and Hexagonal Close Packed (HCP) phases with the predictions from the models. For the lattice constants, the SG-NNP is either the best performing model between the two machine-learned models or the best model amongst all the potentials. The same is true for the formation energies of the diamond, FCC and HCP phases. The SG-NNP model predicts the most accurate lattice constant of the FCC phase, cohesive energy of the BCC phase and formation energy of the diamond phase. The MEAM potential predicts the most accurate BCC and diamond phase lattice constants and the formation energies of FCC and HCP phases. This is surprising because MEAM potential was fitted without diamond, FCC and HCP phase structures in its training database.

Table 5 Lattice Constants (a, c), Cohesive ($E_c$) and Formation ($E_f$) energies of four phases of Molybdenum. Values are either calculated from DFT, interatomic potential or measured from experiments. Values for interatomic potentials are calculated in this work.

|  |  | DFT | SG-NNP | SNAP | EAM | MEAM | Morse | Experiment |
|---|---|---|---|---|---|---|---|---|
| **BCC** | a (Å) | 3.168[a] | 3.163 | 3.160 | 3.147 | 3.167 | 3.141 | 3.150[b] |
|  | $E_c$ (eV) | 6.280[e] | 6.519 | 5.130 | 6.820[f] | 6.817 | 6.754 | 6.820[g] |
| **Diamond** | a (Å) | 5.696[d] | 5.862 | 5.384 | 6.290 | 5.556 | 6.242 |  |
|  | $E_c$ (eV) |  | 4.308 | 4.337 | 4.427 | 4.416 | 2.658 |  |
|  | $E_f$ (eV) | 2.237[h] | 2.211 | 0.793 | 2.393 | 2.401 | 4.186[c] |  |
| **FCC** | a (Å) | 4.012[a] | 4.021 | 4.042 | 3.995 | 3.931 | 3.952 |  |
|  | $E_c$ (eV) |  | 6.287 | 4.835 | 6.746 | 6.426 | 6.844 |  |
|  | $E_f$ (eV) | 0.428[h] | 0.232 | 0.295 | 0.074 | 0.391 | 0.000[c] |  |
| **HCP** | a (Å) | 2.819[a] | 2.834 | 2.869 | 2.825 | 2.743 | 2.795 |  |
|  | c (Å) | 4.583[a] | 4.668 | 4.672 | 4.613 | 4.692 | 4.563 |  |
|  | $E_c$ (eV) |  | 6.280 | 4.917 | 6.746 | 6.403 | 6.844 |  |

| | $E_f$ (eV) | 0.448[h] | 0.239 | 0.214 | 0.074 | 0.415 | 0.000 [c] |

### 2.4.4 Lattice Dynamics

We compare the phonon dispersion curves and phonon density of states of the potentials against experimental data from[47]. Qualitatively, Morse and EAM potentials do not fit the phonon dispersion datapoints as well as the other potentials. The phonon dispersion curves by the SNAP potential show deviations from the experimental throughout the reciprocal directions plotted. For the SG-NNP, notably, the fit along the P-Γ path is almost exact. This is the same for the MEAM. However, it shows discrepancies comparable to the SG-NNP potential for the other directions.

We focus on the various features in the experimental dispersion curves that are not captured by the MEAM potential. The deep local minimum at H is only captured by the SNAP potential though it predicts a crossing of the non-degenerate transverse branch along Γ-N that was not measured in the experiment. The overshooting and retracing of the longitudinal curve over the transverse curve along H-P is only captured by the SG-NNP model.

(a)
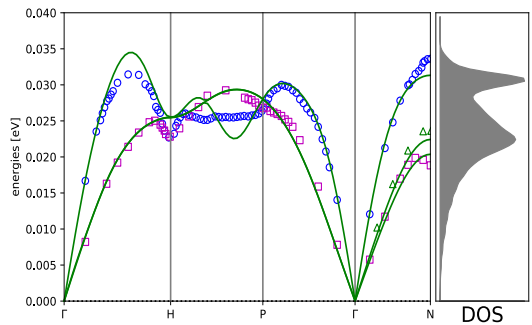

(b)
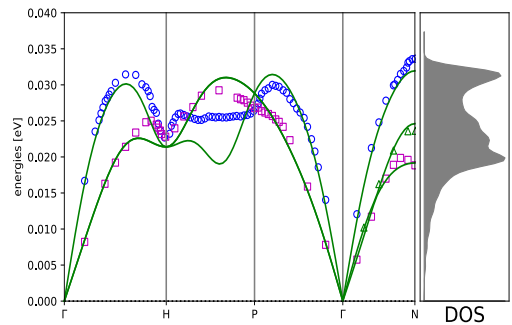

(c)
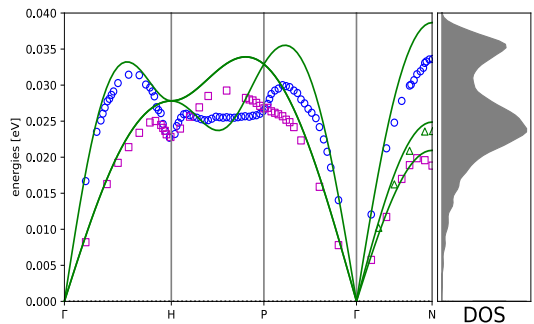

(d)
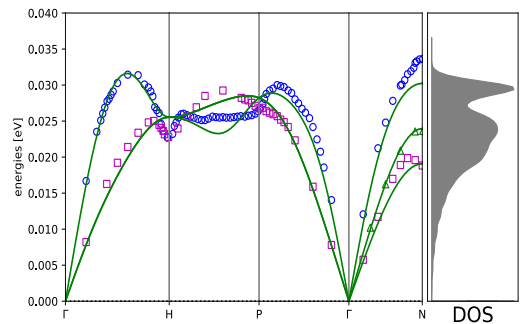

Figure 3 Phonon dispersion curves for all the potentials. The phonon dispersion curve of (a) Neural Network model (b) SNAP (c) EAM and (d) MEAM are shown here in each of the figures as solid green lines. The corresponding phonon density of state (DOS) is also plotted on the right side of each graph. The magenta squares and green triangles represent the experimental dispersion curve of the transverse acoustic modes while the blue circles represent the longitudinal acoustic modes. Morse potential's phonon dispersion is shown in Supplementary Figure 10.

## 2.5 Comparison of Forces and Total Energy Prediction for diverse single and multicomponent Materials

For all materials considered in this section, SG-NNP has the lowest number of dimensions amongst all MLIPs – 22% of SNAP's in the case of NiMo and 12% of HDNNP's for Li. We note at the outset that MLIPs generally have lower errors than EAM, MEAM and Coulomb-Buckingham. Specialized long-range SNAP potential for $Li_3N$, eSNAP, has the lowest errors. MTP has the lowest errors for NbMoTaW, though its dimensionality, expected to be large, was not reported.

We compare SG-NNP with SNAP and HDNNP. Firstly, we focus on elemental materials. SG-NNP predicts equally good or better forces than SNAP. Most of the energies predicted are better than the other potentials by at most 95% (Ge). For the cases where the errors are larger, SG-NNP is only bigger by at most 5% (Ni). Compared with HDNNP, SG-NNP has errors in total energy which are smaller and bigger by at most 174% (Cu) and 16% (Si), respectively. For both cases, we note that SG-NNPs' errors are asymmetrically smaller than bigger. Now, we focus on multicomponent materials and SNAP. SG-NNPs' errors are smaller and bigger by at most 88% (NiMo) and 1% ($Li_3N$), respectively. Again, an asymmetry towards smaller errors for SG-NNP is observed.

Table 6 Dimensionality and error statistics of forces and total energies of various interatomic potentials: SG-NNP, HDNNP, SNAP, eSNAP, MTP, EAM, MEAM and Couloub-Buckingham.

| | Number of Dimensions | | | | | | | | |
|---|---|---|---|---|---|---|---|---|---|
| | Elemental[a] | | | | | | Binary Alloy[b] | | Quaternary Alloy[c] |
| | Ni | Cu | Li | Mo | Si | Ge | NiMo | $Li_3N$ | NbMoTaW |
| **SG-NNP** | 14 | 14 | 14 | 14 | 14 | 14 | 14 | 60 | 60 |
| **SNAP** | 56 | 56 | 56 | 31 | 56 | 56 | 62 | 62 | 124 |
| **HDNNP** | 27 | 27 | 117 | 39 | 27 | 26 | | | |
| **eSNAP** | | | | | | | | 62 | |
| | Test Error of Total Energy (meV/Atom) | | | | | | | | |
| | Ni | Cu | Li | Mo | Si | Ge | NiMo | $Li_3N$ | NbMoTaW |
| **SG-NNP** | 0.82 | 0.42 | 0.81 | 3.93 | 6.63 | 3.69 | 12.00 | 2.32 | 5.03 |
| **SNAP** | 0.78 | 0.53 | 1.03 | 5.31 | 6.89 | 7.19 | 22.50 | 2.30 | 5.10 |
| **HDNNP** | 1.52 | 1.15 | 0.75 | 3.52 | 5.60 | 4.99 | | | |
| **eSNAP** | | | | | | | | 0.82 | |
| **MTP** | | | | | | | | | 4.30 |
| **EAM** | 6.32 | 5.77 | 152.03 | 40.33 | | | 117.2 | | |
| MEAM | 13.12 | 7.32 | | 23.74 | 79.03 | | | | |
| Coulomb-Buckingham | | | | | | | | 22.00 | |
| **Test Error of Forces (eV/Å)** | | | | | | | | | |
| | Ni | Cu | Li | Mo | Si | Ge | NiMo | Li$_3$N | NbMoTaW |
| SG-NNP | 0.03 | 0.02 | 0.02 | 0.14 | 0.11 | 0.09 | 0.13 | 0.11 | 0.08 |
| SNAP | 0.03 | 0.04 | 0.03 | 0.20 | 0.22 | 0.13 | 0.23 | 0.15 | 0.14 |
| HDNNP | 0.03 | 0.03 | 0.02 | 0.12 | 0.11 | 0.09 | | | |
| eSNAP | | | | | | | | 0.04 | |
| MTP | | | | | | | | | 0.06 |
| EAM | 0.06 | 0.07 | 0.10 | 0.29 | | | 0.26 | | |
| MEAM | 0.13 | 0.11 | | 0.13 | 0.23 | | | | |
| Coulomb-Buckingham | | | | | | | | 0.48 | |
| **Error of SG-NNP compared with SNAP** | | | | | | | | | |
| % Difference in Energy Error | 0.05 | -0.26 | -0.27 | -0.35 | -0.04 | -0.95 | -0.88 | 0.01 | -0.01 |
| % Difference in Forces Error | 0.00 | -1.00 | -0.50 | -0.43 | -1.00 | -0.44 | -0.77 | -0.36 | -0.75 |
| **Error of SG-NNP compared with HDNNP** | | | | | | | | | |
| % Difference in Energy Error | -0.85 | -1.74 | 0.07 | 0.10 | 0.16 | -0.35 | | | |
| % Difference in Forces Error | 0.00 | -0.50 | 0.00 | 0.14 | 0.00 | 0.00 | | | |

[a,b,c]Results for SG-NNP are from this work.
[a]Results for all SNAP, HDNNP, EAM and MEAM are from Reference [26].
[b]Results for NiMo (SNAP and EAM) and Li$_3$N (SNAP, eSNAP and Coulomb-Buckingham) are from References [25] and [48], respectively.
[c]Results for SNAP and MTP are from References [7] and [49], respectively.

## 3. Methods

### 3.1 Data

The Mo DFT dataset used to fit the SG-NNP includes total energies and forces of Molybdenum atoms in various configurations. The configurations include grain boundaries, surface slab, vacancies containing structures, and elastically deformed structures of those. They also include snapshots of ab initio molecular dynamics simulations at 300, 3000, and 6000K.

A summary of the datasets used to fit the other Mo potentials is given in Supplementary Table 18.

We refer the reader to the reference for the dataset used for Ni, Cu, Li, Mo, Si, Ge, NiMo, (Ref. [26]), Li$_3$N (Ref. [48]) and NbMoTaW (Ref. [7]).

### 3.2 Species-separated Gaussian Representations

The shift factor construction is crucial to preserving the representation power of RDF and ADF because it preserves the meaning of the height of the δ as a count of the number of atoms or pairs of atoms. Moreover, it spreads out the different types of information to be projected onto different Gaussians, which potentially makes the potential an easier function to learn. Crucially, it limits the number of Gaussian Coefficients needed to describe the distributional space by reuse – a Gaussian kernel could have a large width when its mean is positioned at a region of space where it tends to have little information, to pick up information from other busier parts of the space and help with the prediction.

Other constructions have been proposed by various authors for multiple species with similar RDF and ADF representations. Some involve having a different set of descriptors for every pair of species[34,50] while others involve overloading the heights of the δ with information about the species or species-pairs represented[51]. Both approaches are problematic in their own ways. The former way expands the number of descriptors that have to be used quadratically with the number of species. Our method, in principle, scales with just the number of degrees of freedom in the cutoff radius. We avoid the latter way because it adds informational degeneracy that may affect the representational quality of the RDF and ADF, especially when the number of species is large.

### 3.3 Loss Function and Performance Metric for Investigation of Molybdenum

The loss function, L, we use in our study is defined as such,

$$L = \frac{1}{N_E S_E} \sum_i^{N_E} (E_{i,p} - E_{i,t})^2 + \frac{\beta}{N_F S_F} \sum_i^{N_F} |\vec{F}_{i,p} - \vec{F}_{i,t}|^2 + \lambda \sum_{i,j} W_{i,j}^2 \qquad (8)$$

where $N_E$ is the number of energies, $N_F$ is the number of forces, $E_{i,p}$ is the predicted total energy, $E_{i,t}$ is the target total energy, $\vec{F}_{i,p}$ is the predicted force, $\vec{F}_{i,t}$ is the target force, $W_{i,j}^2$ is the parameters of the model (i.e. weights and biases), $\beta$ is the force beta and $\lambda$ is the L2 regularization constant. We emphasize that $E_{i,p}$ and $E_{i,t}$ refer to total energy of the configuration and not the atomic energies, which are not easily accessible from DFT and so not included in the dataset. Note that we do not include the DFT stresses in the loss as the authors of the Mo SNAP potential [33] did because we believe that getting the accurate forces will naturally lead to accurate stresses.

The performance metric, P, we use in our study is,

$$P = \frac{1}{N_E S_E} \sum_i^{N_E} (E_{i,p} - E_{i,t})^2 + \frac{1}{N_F S_F} \sum_i^{N_F} |\vec{F}_{i,p} - \vec{F}_{i,t}|^2 \qquad (9)$$

where the constants are the same as the ones in loss *L*.

We choose $S_E$ and $S_F$ to have values that are similar to errors reported in [33]. The loss function, L, is used to guide the SG-NNP parameter optimization. The performance metric, P, is used to guide the Gaussian hyperparameter optimization. Note, the larger the performance metric value, the worse the performance.

### 3.4 Parallelization Method

To speed up the whole process, we introduce parallelization at the generation of AGC and RGC and the training of the model. Both steps can be done over multiple nodes, the first using Dask[27] and the second using in-built multi-nodal support from PyTorch [52].

### 3.5 Training Details

#### 3.5.1 Investigation of First-order Optimizer

The cut-off radius, $r_c$, was set at 4Å and five RGCs and five AGCs were employed. A mini-batch is defined to be the total energy and forces on a structure. Unless stated otherwise, we use a mini-batch size of 3, a Gaussian mean search range of [0,1] and a Gaussian width search range of [0.2,1], both in units of the full interval, and the first-order optimizer RMSprop, as it was shown to perform well in many other machine learning cases.

We follow the best practice of using a cyclic learning rate[53], which encourages the model to escape from local minima. We have determined it to be superior to a step decay of the learning rate. In our implementation, we first determined the rate needed for convergence and the initial learning rate in preliminary runs. We then impose a triangular cyclic learning rate schedule. The number of mini-batches to reach the maximum of a cycle is set to 1 and that to descend down towards the minimum is 2700. After each cycle, the maximum learning rate is decreased by a factor of 2. Training was done over 400 epochs and training loss was deemed to be sufficiently converged.

In the two-stage optimization process, a total of 24 independent runs with different Gaussian hyperparameters were done for each stage. We use the whole dataset and a single fold. A cut-off radius of 4 Å results in configurations which include the second-nearest neighbour for the Mo dataset. We use the RMSprop optimizer after a comprehensive test of many optimizers (see results in Supplementary Discussion 2.1.3).

#### 3.5.2 Investigation of Second-order Optimizer

#### 3.5.2.1 Investigation of Molybdenum

We employ a full-connected feedforward NN architecture that has at least two hidden layers. All nodes in hidden layers except the last will have non-linearity of hyperbolic tangent imposed on their outputs. The last layer of the NN will effectively be a dense linear combination of the second-to-last layer of outputs. This is to ensure that each of the outputs from the hyperbolic tangent can be scaled and shifted from the range of [-1,1] of the hyperbolic tangent function to potentially any value, thus giving a greater richness of values to sum up to the atomic energies. We employ the same number of nodes across all hidden layers.

Our SG-NNP training code is implemented using the pyTorch[52] package. In this work, we focus on aspects that are particular to our method.

A second-order LBFGS optimizer[54] with line search and strong Wolfe conditions were used in the training process. Extra care was taken to make sure that the normalization of the descriptors was done separately in each fold to prevent data leakage across folds that may bias the results. In Figure 1 (b), the blue lines are the first-stage runs with hyperparameters of Gaussian coefficients being randomly spaced out in the hyperparameter space, within a user-specified range, using a Latin hypercube methodology[55]. The hyperparameters that were optimized in the specific run are listed in Supplementary Table 19.

### 3.6 Implementation in OpenKIM framework

We implemented the Neural Network potential in the OpenKIM framework[40]. The model can be used with many major molecular dynamics code. We refer the reader to OpenKIM's website [28] for more details. See Supplementary Discussion 3 for the verification checks passed by our potentials.

### 3.7 Mo Models' fitting methods and characteristics

The fitting method employed to generate the potential also affects the quality of its predictions. Supplementary Table 19 is a comparison of these different methods. Chief amongst all considerations is the reported fit quality. SNAP and SG-NNP models have similar absolute MAE in energies and forces. MEAM reported its errors as 11% in force magnitude and 6 degrees in direction when the MEAM force vectors were compared to DFT's. Both of the potentials and MEAM were fitted on the DFT energies and forces. While EAM fits on derived energies from DFT calculations, it does not fit directly on DFT total energies or forces.

All potentials, except Morse, employed a two-stage fitting process where some hyperparameters were first optimized before the rest of the parameters. EAM fits some of the properties in its first-stage exactly before fitting the rest inexactly. Morse fits its experimental values exactly. The SG-NNP potential only employed forces and energies in the fitting process, while SNAP also uses the elastic constants to fit hyperparameters in the first stage and stresses in the second stage.

### 3.8 Response Surface Methodology and Latin Hypercube

Using a response surface methodology[56] and cubic radial basis functions for interpolation, we obtain optimized hyperparameter values from the first-stage runs that could potentially maximize the test performance. We emphasize here that we are optimizing the average cross-validation errors and not the training errors.

### 3.9 Calculations

### 3.9.1 Lattice Dynamics

We use the phonon package which implements the small displacement method[57] in ASE[58], a Python library for atomistic simulations.

### 3.9.1 Neural Network Training

We perform training runs of our Neural Network model written in Pytorch 2.0[59] on a GPU workstation with Nvidia RTXA6000 card.

**Data Availability**

Datasets for Ni, Cu, Li, Mo, Si, Ge, NiMo, Li$_3$N and NbMoTaW are available at https://github.com/materialsvirtuallab/mlearn/tree/master/data and https://github.com/materialsvirtuallab/snap.

**Code Availability**

Code for the preprocessing of the raw data and the code for training on processed data is available on reasonable request from the authors.

**Acknowledgements**

We gratefully acknowledge the invaluable discussions with Professor Mark Asta, whose insights and expertise significantly contributed to the development of this research. We acknowledge the code for the differential displacement map plotter from Dr. Sheng Yin. We also thank the Savio Supercomputer at the University of California, Berkeley, for providing the computational resources necessary for some of our simulations, model training and inference. This study was supported by the Accelerated Materials Development for Manufacturing Program at A*STAR via the AME Programmatic Fund by the Agency for Science, Technology and Research, Singapore under Grant No. A1898b0043.

**Author contributions**

JWY conceptualized the project and methodology, curated the data, developed the software, validated and visualized the results. JWY wrote the original draft of the manuscript, reviewed and edited subsequent versions. BJZ developed the software, investigated, visualized and validated the results. JS was involved with the project administration, investigation, funding acquisition, resources, review & editing of the manuscript and supervision of the project.